\documentclass[a4paper,11pt]{article}
\usepackage{pos}
\usepackage{makecell}
\usepackage{float}
\usepackage{hyperref}

\title{Impact of Plasma Instabilities and of the Intergalactic Magnetic Field on Blazar-Induced Electromagnetic Cascades}
\ShortTitle{Impact of Plasma Instabilities and Intergalactic Magnetic Field on Cascades}

\author*[a]{Suman Dey}
\author[a]{G\"{u}nter Sigl}

\affiliation[a]{II. Institut f\"{u}r Theoretische Physik, Universit\"{a}t Hamburg,\\
  Luruper Chaussee 149, Hamburg, 22761, Germany.}

\emailAdd{suman.dey@desy.de}
\emailAdd{guenter.sigl@desy.de}

\abstract{Intergalactic weak magnetic fields can have non-negligible effects on the electromagnetic cascades induced by blazar gamma-ray emission. Secondary electrons and positrons are produced by primary gamma rays of energies ~TeV and can be magnetically deflected out of the line of sight to the observer. However, these leptons can perturb the background intergalactic medium (IGM), resulting in the growth of plasma instabilities, which can also influence the electromagnetic cascade. The resulting gamma-ray spectrum, observable in the GeV-TeV energy range, can bear imprints of these two competing phenomena: deflection by the intergalactic magnetic field and plasma instability cooling. We present the results of numerical simulations that incorporate the combined impact of these two processes on the propagated gamma-ray spectrum of the blazar 1ES 0229+200.}

\ConferenceLogo{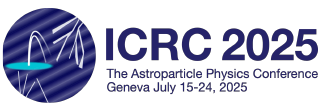}

\FullConference{39th International Cosmic Ray Conference (ICRC2025)\\
 15–24 July 2025\\
Geneva, Switzerland\\}
\begin{document}
\maketitle

\section{Introduction}
A blazar is a specific type of active galactic nuclei (AGN) that produces a relativistic jet of TeV gamma rays that is oriented along the line of sight to Earth. The TeV photons propagate through the intergalactic medium (IGM) and undergo interactions with the extragalactic background light (EBL) photons ($\gamma_{\text{EBL}}$) \cite{Hauser:2001xs}, leading to the production of relativistic electron-positron pairs via the process $\gamma_{\text{TeV}} + \gamma_{\text{EBL}} \rightarrow e^+ + e^-$. The higher-order processes, such as double pair production (DPP) and triple pair production (TPP), are suppressed at energies below $\sim100~\text{TeV}$ due to their comparatively larger interaction lengths (see \cite{Heiter:2017cev}). The pairs subsequently upscatter cosmic microwave background (CMB) photons \cite{gould1967opacity, blumenthal1970bremsstrahlung} through the inverse Compton (IC) scattering process $e^{\pm} + \gamma_{\text{CMB}} \rightarrow e^{\pm} + \gamma_{\text{GeV}}$, resulting in the production of an electromagnetic cascade of GeV photons. However, the observations from experiments such as Fermi-LAT and ground-based Cherenkov telescopes (e.g., H.E.S.S., MAGIC, VERITAS) \cite{HESS:2023zwb, Romoli:2017nrp, SilvaBatista:2023jcr, MAGIC:2022piy} show a discrepancy between the predicted and observed secondary GeV photon fluxes \cite{aharonian2001tev, neronov2009sensitivity, neronov2010evidence}. One possible explanation for the missing GeV cascade emission is the deflection of electron-positron by intergalactic magnetic fields (IGMF), resulting in a smaller gamma ray flux within the field of view of the detector. An alternative explanation for the missing GeV flux is the energy loss of electrons/positrons through plasma instabilities. The interactions between the relativistic blazar-induced pair beam and the IGM plasma can give rise to electrostatic instabilities. Plasma instabilities offer an additional channel for energy dissipation, potentially competing with IC cooling. The efficiency and relevance of these instabilities remain subjects of active theoretical and numerical investigation \cite{Broderick:2011av, Miniati:2012ge, schlickeiser2012plasma, schlickeiser2013plasma, Sironi:2013qfa, vafin2018electrostatic, AlvesBatista:2019ipr, Alawashra:2024fsz, Dey:2025wdj}. In this work, we study the lower bound on the IGMF in the presence of plasma instability. We employ the publicly available Monte Carlo simulation framework \texttt{CRPropa3.2} \cite{AlvesBatista:2022vem}\footnote{\href{https://crpropa.desy.de/}{https://crpropa.desy.de/}} to model the propagation of electromagnetic particles (electrons, positrons, and photons).

When a TeV blazar emits gamma rays continuously, and the emission period is much longer than the typical delay time for cascade photons, the effect of the delay is "washed out," and the observer sees the total, steady-state emission. Given that the blazar 1ES 0229+200 shows no significant rapid variability, we focus our analysis on the angular extension of the observed emission rather than on the time delay analysis approach used in \cite{MAGIC:2022piy, Vovk:2023qfk}.

\section{Energy loss due to plasma instability}
The energy loss of pair beams due to the linear growth of instability can be quenched by the non-linear feedback of the instability, which further corresponds to the saturation of instability. The non-linear feedback of the instability causes a negligible angular broadening of the order of $\sim 10^{-4}$ rad. Thus, we consider that our one-dimensional approximation of the energy loss term remains valid in the regime where the electrons/positrons lose less than $\sim 2\%$ of their energy before the quenching of instability occurs between successive IC scatterings \cite{Alawashra:2024fsz, Dey:2025wdj}. We define the generalized energy-loss term (a similar approach as in \cite{Castro:2024ooo}) induced by plasma instabilities on electron-positron pairs from well-known models \cite{Broderick:2011av, Miniati:2012ge, schlickeiser2012plasma, schlickeiser2013plasma, Sironi:2013qfa, vafin2018electrostatic}. These plasma instability models are based on analytical calculations and kinetic simulations. We have studied the fractional energy loss due to instability within the interaction length of one IC scattering.
%, rather than on the relationship between the energy loss rate and instability growth rate. 
%Figure \ref{fig:chart} summarizes the generalized energy-loss term used in our study, based on well-known plasma instability models. 
%\begin{figure}[h]
%    \centering
%    \includegraphics[width=1.0\linewidth]{ICRC2025_template/chart-3.pdf}
%    \caption{Schematic overview of well-known plasma instability models, where the energy loss rate is generalized using three parameters: the normalization electron energy $\tilde{E_{e}}$, the instability length scale $\lambda_{0}$, and the power-law index $\alpha$.}
%    \label{fig:chart}
%\end{figure}
If $E_{e}(x)$ is the energy of an electron at a distance $x$ from the point of injection, then the energy loss of the electron within an IC interaction length, $\lambda_{\text{IC}}$, is $\Delta E_{e}(\lambda_{\text{IC}}) = E_{e}(x_0) - E_{e}(\lambda_{\text{IC}})$, where $x_0$ is the injection point. We consider the regime where the fractional energy loss due to instability in an IC interaction length, $\Delta E_{e}(\lambda_{\text{IC}})/E_{e}$, is less than 1, implying that electrons/positrons lose only a small fraction of their energy through instability within an IC interaction length. The energy loss rate due to instability is defined as $E_{e}^{-1}(dE_{e}/dt)= c \lambda_{0}^{-1} (E_e/\tilde{E_e})^{-\alpha}~\text{sec}^{-1}$, where we conduct a parametric study by varying the instability length scale, $\lambda_{0}$ and power-law index, $\alpha$, while the normalization electron energy is set to $\tilde{E_{e}} = 1 ~\text{TeV}$. A more detailed study of this plasma instability model can be found in \cite{Castro:2024ooo}.

\section{Turbulent magnetic field setup}
We implement a homogeneous turbulent magnetic field with a Kolmogorov spectrum, characterized by the Kolmogorov index $\mathscr{k} = 5/3$. The coherence length $\lambda_c$ is determined by the maximum and minimum length values $\lambda_{\max}$ and $\lambda_{\min}$, and is given by the relation \cite{Dundovic:2017vsz}
\begin{equation}
\lambda_c = \frac{1}{2\lambda_{\max}} \cdot \frac{\mathscr{k} - 1}{\mathscr{k}} \cdot \frac{1 - (\lambda_{\min}/\lambda_{\max})^\mathscr{k}}{1 - (\lambda_{\min}/\lambda_{\max})^{\mathscr{k} - 1}}.
\end{equation}
Since $\lambda_{\max}$ and $\lambda_{\min}$ are the actual inputs to the simulation, one needs to numerically solve this equation to match a target coherence length $\lambda_c$. However, the discretized nature of the magnetic field grid in CRPropa imposes constraints on the allowable turbulence scales: for a grid with $N$ cells of size $\Delta x$, the turbulent field is properly contained only if $\lambda_{\max} < N\Delta x$ and $\lambda_{\min} > 2\Delta x$. This restricts the range of usable $\lambda_c$ values. Consequently, the choice of $\lambda_c$ directly impacts the possible magnetic field strength, which is constrained by $B \simeq E_{e}\delta/(|q_{e}|c\sqrt{L\lambda_c})$, where $L$ is the distance travelled by the electron and $\delta$ is the deflection angle of the electron by IGMF, relative to the direction of the primary gamma ray \cite{Sigl:2016yun}. For our configuration with $N = 256$, the parameters $\lambda_{\min}$ and $\lambda_{\max}$ are selected in accordance with the grid limitations to generate a physically consistent turbulent field.
%$B \lesssim \frac{E_{e} \delta}{|q_{e}|c}\frac{1}{\sqrt{d \lambda_c}}$ --- before undergoing an IC scattering.

\section{Cascade simulations of Blazar 1ES 0229+200}
We use the publicly available, 3-dimensional Monte Carlo simulation framework \texttt{CRPropa3.2} \cite{AlvesBatista:2022vem} for electromagnetic cascade simulations. In our simulation, we model the 1ES 0229+200 blazar jet as a narrow emission cone of $10^{\circ}$ following a von Mises–Fisher distribution \cite{Jasche:2019sog}, with the emission direction oriented along the positive $x$-axis and a concentration parameter $\kappa \simeq 200$. The high $\kappa$ value ensures a strongly collimated particle distribution around the jet axis, consistent with the small opening angles expected from relativistic blazar jets. We model the observer as a sphere centered at the origin, with a radius equal to the distance between the source and the observer, $d=580$ Mpc at redshift $z \sim 0.14$. We consider the pair production, IC scattering, DPP, and TPP interaction modules that are already available in \texttt{CRPropa3.2}, taking into account the cosmic microwave background (CMB) and the infrared background (IRB) model from \cite{franceschini2008extragalactic} as ambient photon fields. The photon spectrum $J(E; \lambda_{0}, \alpha; \theta_{\text{FoV}}, \lambda_{c}, B)$ observed at Earth can be obtained by folding the cascade signal $G(E_{0}, E;\lambda_{0}, \alpha; \theta_{\text{FoV}}, \lambda_{c}, B)$ with the initial injection spectrum $J_{0}(E_{0})$, expressed as \cite{MAGIC:2022piy},
\begin{equation}
    %\begin{split}
        J(E; \lambda_{0}, \alpha; \theta_{\text{FoV}}, \lambda_{c}, B) = \int_{E_{0}\geq E} G(E_{0}, E;\lambda_{0}, \alpha; \theta_{\text{FoV}}, \lambda_{c}, B)\cdot J_{0}(E_{0})~dE_{0},
    %\end{split}
    \label{eq:reconstruction-spectrum}
\end{equation}
We consider an injected energy spectrum given by a power-law function with an exponential cutoff, expressed as,
\begin{equation}
    J_{0}(E_{0}) = A \left(\frac{E_{0}}{\text{TeV}}\right)^{-\beta} \text{exp}\left(-\frac{E_{0}}{E_{\text{cut}}}\right),
    \label{eq:power-law}
\end{equation}
where $E_{0}$ is the energy of the primary photon, $\beta$ represents the source spectral index, $E_{\text{cut}}$ is the high-energy cutoff, and $A$ is a normalization constant. We simulate $10^{4}$ primary photons with a propagation step size of $10^{14}$ cm.

The momentum transfer between the primary TeV photon, the EBL photon, and the produced electron-positron pair is governed by relativistic kinematics. In the IGM rest frame (lab frame), the pair production process occurs when a primary TeV gamma-ray, with four-momentum $P_{0} = E_{0}(1, \mathbf{\hat{p}_{0}})$, interacts with a low-energy EBL photon having the four-momentum $P_{1} = E_{\text{EBL}}(1, \mathbf{\hat{p}_{1}})$, where $\mathbf{\hat{p}_{0}}$ and $\mathbf{\hat{p}_{1}}$ are unit vectors along their momentum. The four-momentum of the resulting electron and positron is given by $P_{\pm} = (E_{e\pm}, \mathbf{q_{\pm}})$ \cite{Schlickeiser:2012hn}. It is worth noting that the initial angle between the momentum of incident photon (or the beam axis direction) and the relativistic momentum of the pair-produced electron/positron beam with Lorentz factor $\Gamma$ (typically in the range $10^{5}$–$10^{7}$) is approximately $\sim \Gamma^{-1}$, and thus negligible during the pair production process \cite{Perry:2021rgv}. Before up-scattering CMB photons to high energies ($\sim$ GeV) through inverse Compton scattering, the electrons (and positrons) are deflected by the IGMF and experience energy loss due to plasma instability. The deflection angle of the electron by IGMF, relative to the direction of the primary gamma ray, is given by $\delta = \text{cos}^{-1}(\mathbf{\hat{p}_{0}\cdot\hat{q}_{\pm}^{\prime}})$, where $\mathbf{\hat{p}_{0}}$ and $\mathbf{\hat{q}_{\pm}^{\prime}}$, i.e., the unit vector of momentum of electron (or positron) after deflection can be directly obtained from the simulation output. %The schematic representation shows that 
The IC up-scattered secondary photons with energy $E_{\gamma}$ arrive within an angle $\theta_{\text{obs}}$, which can be expressed in terms of the mean free path ($\lambda_{\gamma}$) of $\gamma$-rays of initial energy $E_0$, the distance between the source to the observer in the comoving frame, $d$, and $\delta$ \cite{Neronov:2009gh},
\begin{equation}
    \text{sin}\left(\theta_{\text{obs}}\right) = \frac{\lambda_{\gamma}\left(E_{0}\right)}{d} \text{sin }\delta.
    \label{extended-angle}
\end{equation}
From the equation above, we estimate $\theta_{\text{obs}}$, then define the observer field of view angle $\theta_{\text{FoV}}$ as an independent parameter and measure the cascade-induced "glow" within this field of view. The angular resolution or point-spread function (PSF) of the Fermi-LAT is energy-dependent, with a 68\% containment angle of about $0.1^{\circ}$ at $30$ GeV \cite{atwood2009large, Fermi-LAT:2013jgq}\footnote{\href{https://www.slac.stanford.edu/exp/glast/groups/canda/lat_Performance.htm}{https://www.slac.stanford.edu/exp/glast/groups/canda/lat\_Performance.htm}}. We choose representative values of $\theta_{\text{FoV}}=0.2^{\circ}$ and $4.5^{\circ}$, consistent with the observational requirement $\theta_{\text{psf}} < \theta_{\text{obs}} < \theta_{\text{FoV}}$ \cite{AlvesBatista:2021sln}. For comparison, the Large-Sized Telescope (LST) of the Cherenkov Telescope Array Observatory (CTAO) offers a total field of view of about $4.3^{\circ}$ and sensitivity in the $20-150$ GeV range\footnote{\href{https://www.ctao.org/emission-to-discovery/telescopes/lst/}{https://www.ctao.org/emission-to-discovery/telescopes/lst/}}. In future studies, combining upcoming CTA and Fermi-LAT observations could result in a larger effective FoV. We simulate the photon spectrum for different configurations of magnetic field strengths and $\lambda_{c}$ values within these fields of view, taking into account the effects of plasma instabilities. To explore the influence of plasma instabilities on the cascade development and to constrain the intergalactic magnetic field (IGMF), we consider three representative scenarios.
\begin{itemize}
    \item \textit{Case 1: }This scenario assumes the cascade spectrum only from the influence of the IGMF, without any contribution from plasma instabilities;
    \item \textit{Case 2: }In this scenario, plasma instabilities are modeled with $\lambda_0 = 600~ \text{kpc}$, $\alpha = -0.5$, and $\tilde{E_e} = 1.0 ~\text{TeV}$. The extended emission results from the combined effects of both the IGMF and the plasma instability;
    \item \textit{Case 3: }The plasma instability parameters are considered slightly stronger with $\lambda_0 = 120~ \text{kpc}$, $\alpha = -0.5$, and $\tilde{E_e} = 1.0~\text{TeV}$. Similar to the previous case, the extended emission is produced by the simultaneous effects of the IGMF and plasma instabilities.
\end{itemize}
For each case, we simulate the resulting photon spectrum and determine the corresponding IGMF strength and $\lambda_{c}$ that provides the best fit to the observational data.

We define the $\chi^{2}$ test statistic to determine the deviation of the simulated spectrum from the observation \cite{Castro:2024ooo},
\begin{equation}
%\begin{split}
    \chi^{2}(\beta, E_{\text{cut}}; \lambda_{0}, \alpha; \theta_{\text{FoV}}, \lambda_{c}) = \sum\left(\frac{J_{\text{data}}(E_{i})-J_{\text{sim}}(E_{i};\beta, E_{\text{cut}}; \lambda_{0}, \alpha; \theta_{\text{FoV}}, \lambda_{c})}{\sigma_{\text{data},i}}\right)^{2},
 %   \end{split}
    \label{eq:chisq}
\end{equation}
where $J_{\text{data}}(E_{i})$ and $J_{\text{sim}}(E_{i};\beta, E_{\text{cut}}; \lambda_{0}, \alpha; \theta_{\text{FoV}}, \lambda_{c}, B)$ represent the observed and propagated photon spectra at Earth, respectively, evaluated at the same energy $E_{i}$. The summation is carried out over the observed energy bins. We calculate the normalised $\chi^{2}/n_{dof}$ for each set of simulation runs and estimate the corresponding $\chi^{2}_{\text{min}}/n_{dof}$, corresponding to the best fit of the photon spectrum to the observed GeV data. We perform the fit to the observed GeV data because cascade emission primarily dominates the GeV energy range, while the TeV part remains unchanged. 
%We perform the fit to the observed GeV data.
%, which typically leaves the best-fit results largely unaffected. We fit the simulated photon spectra to the observed GeV data because the electromagnetic cascade flux modifies the total spectrum in the GeV region, while the high-energy part remains largely unchanged.  

\section{Modified lower limit of IGMF due to plasma instability}
The relativistic electrons are produced from the TeV photons of energies $0.6\lesssim E_{e}/\text{TeV} \lesssim 5.6$ that interact with the CMB photons and produce the secondary $\gamma-$rays with energies $1\lesssim E_{\gamma}/\text{GeV} \lesssim 100$ \cite{Neronov:2009gh}. These electrons lose energy through IC scattering over a characteristic length scale of $0.1\lesssim l_{\text{IC}}/\text{Mpc}\lesssim0.6$ \cite{Neronov:2009gh}. If the magnetic coherence length is longer than the length scale of IC ($l_{\text{IC}}\ll\lambda_{\text{c}}$), then the deflection angle of the $e^{+}e^{-}$ pairs, $\delta = l_{\text{IC}}/r_{L}$, becomes independent of the coherence length. Here, $r_{L} = E_{e}/(|q_{e}|B)$ is the Larmor radius in a magnetic field $B$. In contrast, in the regime $l_{\text{IC}}\gg\lambda_{\text{c}}$, the deflection follows a random walk, and particles experience random magnetic field orientations, causing an angular spread. In this case, the field geometry becomes turbulent, causing random-walk deflection with a deflection angle, $\delta_{\text{cell}} \simeq \lambda_{\text{c}}/r_{L}$. After $n=l_{\text{IC}}/\lambda_{\text{c}}$ cells, the cumulative effect of these small deflections over $\lambda_{\text{IC}}$ distance leads to an overall deflection angle, $\delta \simeq \delta_{\text{cell}}\sqrt{l_{\text{IC}}/\lambda_{\text{c}}} = |q_{e}|B\sqrt{l_{\text{IC}}\lambda_{\text{c}}}/E_{e}\propto B\sqrt{\lambda_{\text{c}}}$ \cite{Vovk:2023qfk, Korochkin:2020pvg}.
\begin{figure}[ht]
    \centering
    \includegraphics[width=0.338\textwidth]{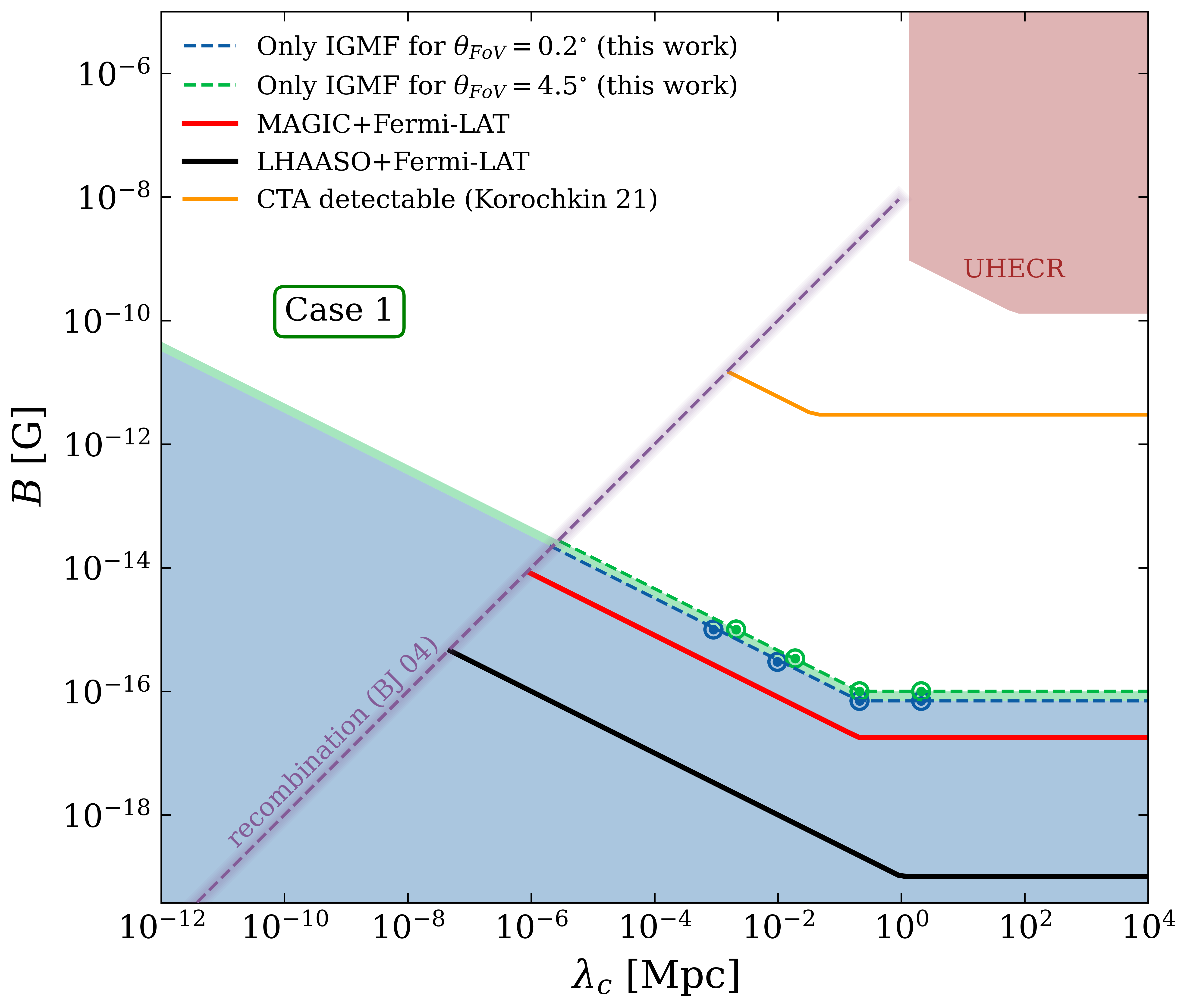}
    \includegraphics[width=0.32\textwidth]{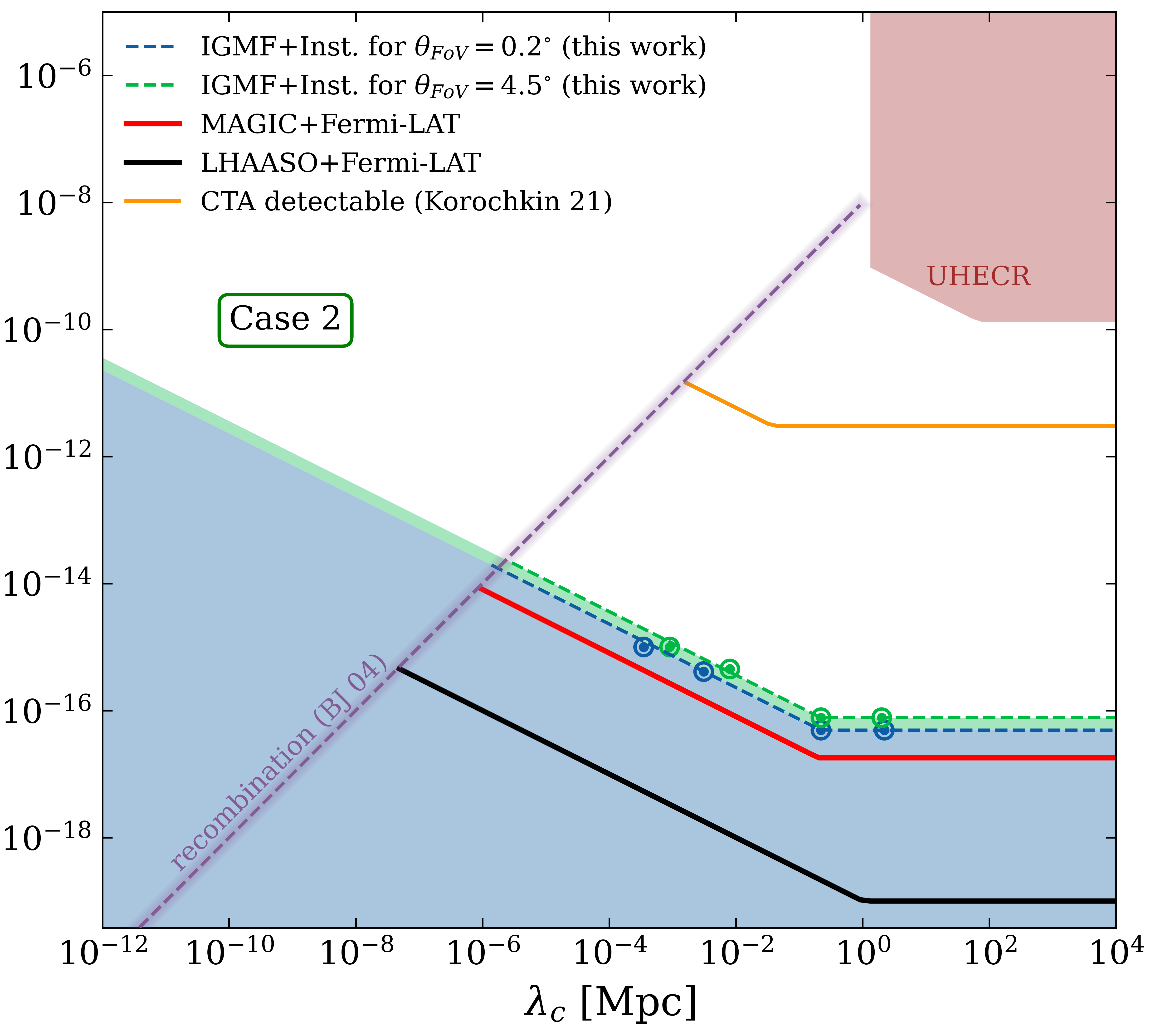}
    \includegraphics[width=0.32\textwidth]{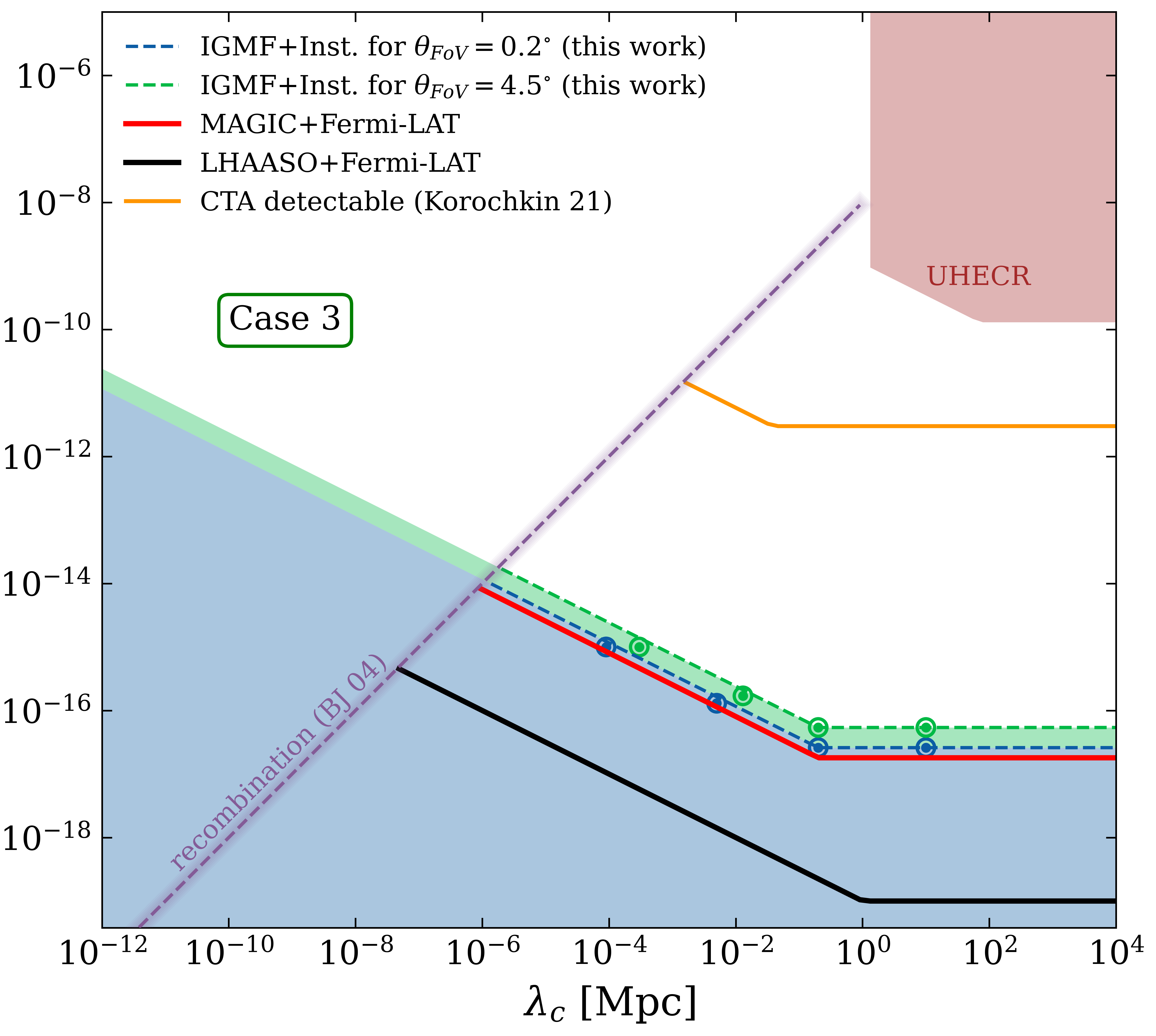}
    \caption{The lower limit of the IGMF due to instability inferred from blazar 1ES 0229+200 is shown in comparison with existing constraints from MAGIC+Fermi-LAT \cite{MAGIC:2022piy} (red line), LHAASO+Fermi-LAT observations derived from GRB 221009A \cite{Vovk:2023qfk} (black line), UHECR observations \cite{Neronov:2021xua} (light brown shaded region), with predictions from cosmological primordial magnetic field evolution models \cite{Banerjee:2004df} (glowing purple dashed line), and IGMF limits that will be detectable by the CTA \cite{Korochkin:2020pvg} (orange line). The green and blue dashed lines define the IGMF limits derived for $\theta_{\text{FoV}}=0.2^{\circ}$ and $4.5^{\circ}$, respectively. The first panel shows the IGMF constraint without considering plasma instabilities. The second and third panels present the modified limits obtained by including the effects of plasma instability for the respective parameter choices.
}
    \label{fig:b-lc}
\end{figure}
Figure \ref{fig:b-lc} shows that in the absence of plasma instabilities, the IGMF lower limit is quite strong, approximately of the order of $B\simeq7\times10^{-17}~\text{G}$, within an observer $\theta_{\text{FoV}}=0.2^{\circ}$. When plasma instabilities are taken into account, the IGMF lower limit weakens to about $B \simeq 2.6 \times 10^{-17}~\text{G}$ within an observational field of view of $\theta_{\text{FoV}} = 0.2^{\circ}$. The IGMF limits for different plasma instability parameters are summarised in Table \ref{tab:summarised}. The red line in the first panel of Figure \ref{fig:b-lc} indicates that the IGMF lower limit appears slightly weaker than our estimation. This discrepancy arises because the MAGIC+Fermi-LAT study adopts a time-delay strategy with a maximum geometrical time delay of $t_{d} = 10$ years. However, the expected $t_{d}$ for $10$ GeV gamma rays, assuming $B \gtrsim 10^{-17}$ G with $\lambda_{c} \gtrsim 0.1~\text{Mpc}$, is always $t_{d} \gtrsim 100$ years, while for 60 GeV gamma rays under the same conditions it is $t_{d} \gtrsim 10$ years \cite{AlvesBatista:2021sln, neronov2009sensitivity}. As a result, the contribution of low-energy gamma rays to cascade production is underestimated, and complete estimation of the extended emission within a realistic time window of a few decades becomes difficult. On the other hand, according to our field-of-view analysis strategy, for $1$ GeV gamma rays, the extended emission has a size of $\theta_{\text{obs}} \simeq 0.02^{\circ}$, assuming $B \gtrsim 10^{-17}$ G and $\lambda_{c} \gtrsim 0.1~\text{Mpc}$. Under our chosen values of $\theta_{\text{FoV}}$, the cascade-induced "glow" remains fully contained within the observer’s field of view.
\begin{table*}[ht]\footnotesize\centering
\caption{The summary of IGMF limits depending on plasma instability parameters.}\label{tab:summarised}
\begin{tabular}{c*{6}{c}r}%{l*{6}{c}r}
 %&\multicolumn{2}{c}{$D_{4h}^1$}&\multicolumn{2}{c}{$D_{4h}^5$}\\
 \hline
 Case & $\theta_{\text{FoV}}=0.2^{\circ}$ & $\theta_{\text{FoV}}=4.5^{\circ}$ & {}\\ \hline 
\makecell{\text{1.}} & $B\gtrsim\begin{cases}
			7.0\times 10^{-17}\text{~G}, & \\
            7.0\times 10^{-17}(\lambda_{\text{c}}/0.21 \text{~Mpc})^{-1/2} \text{~G}, &
		 \end{cases}$ & $\begin{cases}
			1.0\times 10^{-16}\text{~G}, & \lambda_{\text{c}}> 0.21 \text{~Mpc}\\
            1.0\times 10^{-16}(\lambda_{\text{c}}/0.21 \text{~Mpc})^{-1/2} \text{~G}, & \lambda_{\text{c}}< 0.21 \text{~Mpc}
		 \end{cases}$ \\ \hline
\makecell{\text{2.}} & $B\gtrsim\begin{cases}
			4.9\times 10^{-17}\text{~G}, & \\
            4.9\times 10^{-17}(\lambda_{\text{c}}/0.22 \text{~Mpc})^{-1/2} \text{~G}, &
		 \end{cases}$ & $\begin{cases}
			7.7\times 10^{-17}\text{~G}, & \lambda_{\text{c}}> 0.22 \text{~Mpc}\\
            7.7\times 10^{-17}(\lambda_{\text{c}}/0.22 \text{~Mpc})^{-1/2} \text{~G}, & \lambda_{\text{c}}< 0.22 \text{~Mpc}
		 \end{cases}$ \\ \hline
\makecell{\text{3.}} & $B\gtrsim\begin{cases}
			2.6\times 10^{-17}\text{~G}, \\
            2.6\times 10^{-17}(\lambda_{\text{c}}/0.20 \text{~Mpc})^{-1/2} \text{~G},
		 \end{cases}$ & $\begin{cases}
			5.4\times 10^{-17}\text{~G}, & \lambda_{\text{c}}> 0.20 \text{~Mpc}\\
            5.4\times 10^{-17}(\lambda_{\text{c}}/0.20 \text{~Mpc})^{-1/2} \text{~G}, & \lambda_{\text{c}}< 0.20 \text{~Mpc}
		 \end{cases}$ \\ \hline 
%\makecell{\text{No Instability}} & $\text{\textcolor{red}{TO BE FOUND}}$ & $\text{\textcolor{red}{TO BE FOUND}}$ \\
\end{tabular}
\end{table*}

\section{Conclusion}
We perform a parametric study of the energy loss due to plasma instability by simulating the extragalactic propagation of high-energy photons from the blazar 1ES 0229+200 in the presence of IGMF. We minimize the $\chi^{2}_{\text{min}}/n_{dof}$ for different configurations, first considering only the magnetic field and then including two different plasma instability cases, to estimate the lower limit of the IGMF while determining the source parameters $\beta, E_{\text{cut}}$ that best reproduce the observed photon spectrum. 
%We minimize the $\chi^{2}_{\text{min}}/n_{dof}$ to estimate the lower limit of IGMF, for the optimal combination of instability parameters $\alpha$, $\lambda_{0}$ and source parameters $\beta$, $E_{\text{cut}}$ that best reproduce the observed photon spectrum.
The simulation output provides an estimate of the cascade signal as a function of the observer's field of view angle. The scenario with plasma instability parameters $\lambda_0 = 120~ \text{kpc}$, $\alpha = -0.5$, and $\tilde{E} = 1.0~\text{TeV}$ yields an IGMF lower limit estimate of $B\gtrsim2.6\times10^{-17}~\text{G}$, within an observer $\theta_{\text{FoV}}=0.2^{\circ}$ that closely aligns with the lower bound reported by \cite{MAGIC:2022piy} for the same blazar. For this configuration, the Fermi-LAT data yield a minimum $\chi^{2}_\text{min}/n_{dof}=1.802$, corresponding to a source spectral index $\beta = 1.22$ and a high-energy cut-off $E_{\rm cut} = 10.7$ TeV. In this case, the corresponding fractional energy loss within an IC interaction length is $\lesssim 2\%$ for electrons with energies below $4$ TeV and $\lesssim 
1\%$ for electrons in the GeV energy range (as reported by \cite{Castro:2024ooo}). In future work, combining upcoming LST-CTAO and Fermi-LAT observations could yield a larger effective FoV. The findings of this study are especially relevant for both ongoing and upcoming observational strategies.
%, making the findings of this work especially relevant for ongoing observational strategies. %would yield a larger effective FoV, making the findings of this study particularly relevant to current observational strategies.
%In the future, The LST is optimized for low-energy gamma-ray observations in the $20-150$ GeV range and features a wide field of view of approximately 4.3$^\circ$, 

\section{Acknowledgments}
This research is funded by the Deutsche Forschungsgemeinschaft (DFG, German Research Foundation) under Germany’s Excellence Strategy– EXC 2121 “Quantum Universe”– 390833306. The authors gratefully acknowledge the CRPropa simulation framework. The authors acknowledge the HPC facility of the PHYSnet cluster operated at Universit\"{a}t Hamburg, Germany, and Phenod cluster computational resources operated at Deutsches Elektronen-Synchrotron (DESY), Hamburg, Germany. 

\bibliographystyle{apsrev4-2}
\bibliography{ref}% Produces the bibliography via BibTeX.
%\begin{thebibliography}{99}
%\bibitem{...}
%....

%\end{thebibliography}

\end{document}